# Accelerating Structure design and fabrication For KIPT and PAL XFEL

Mi Hou (候汨), Xiang He (贺祥), Pei Shi-lun (裴世伦)

**Abstract**: ANL and the National Science Center "Kharkov Institute of Physics Technology" (NSC KIPT, Kharkov, Ukraine) jointly proposed to design and build a 100MeV/100KW linear accelerator which will be used to drive the neutron source subcritical assembly. Now the linac was almost assembled in KIPT by the team from Institute of High Energy Physics (IHEP, Beijing, China). The design and measurement result of the accelerating system of the linac will be described in this paper.

**Key words**  S-band, accelerating structure, BBU effect, neutron source.

**PACS**  Accelerators, 29.20.-c

## 1 Introduction

A neutron source based on the sub-critical assembly driven by electron linear accelerator was being constructed[1]. This is a project with the participation of NSC KIPT, ANL and IHEP together. For providing a neutron flux of about $10^{13}$ neutron/s, the design of a linear accelerator with average electron beam power of 100 kW/100 MeV is needed[2]. The main specifications of the accelerator are listed in Table 1.

Table 1. NSC KIPT neutron source Parameters

| Parameter | Values | Units |
|---|---|---|
| RF frequency | 2856 | MHz |
| Beam energy | 100 | MeV |
| Beam current | 0.6 | A |
| Beam spread (peak to peak) | 4 | % |
| Emittance (1 s) | $5*10^{-7}$ | m*rad |
| Beam pulse duration | 2.7 | ms |
| RF pulse width | 3 | ms |
| RF repetition rate (max) | 625 | Hz |

To satisfy the optimized physical design, the accelerating system of the KIPT linac consists of a 2856 MHz single cavity prebuncher, a 2856 MHz traveling wave buncher and ten 1.34-meters-long 2856 MHz traveling wave accelerating structures. A diagram illustrating the NSC KIPT linac is shown in Figure 1.

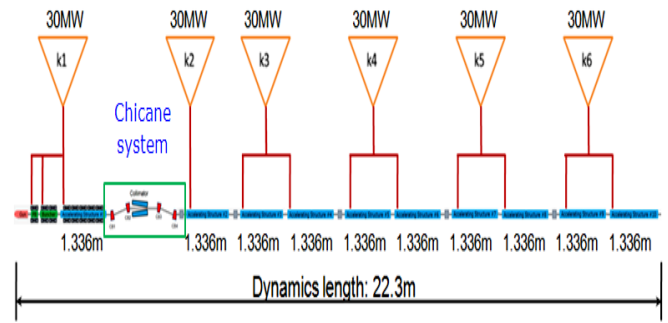

Figure 1. Systemic diagram of the NSC KIPT linac

## 2 Prebuncher

Prebuncher is a re-entrant resonant standing-wave cavity. A single side coupling mechanism is adopted which is different from the conventional design. As a consequence, an eccentric circle structure is adopted to avoid the asymmetry of the E-field along the transverse direction going through the center of the coupling iris. The main parameters of the prebuncher are listed in Table 2.

Table 2. Main parameters of prebuncher

| Parameter | Unit | BEPC | KIPT |
|---|---|---|---|
| Operating frequency | MHz | 2856 | 2856 |
| Unloaded $Q_0$ | | 1000 | 1446 |
| Cavity diameter | mm | 50.8 | 52.62 |
| Nose cone diameter | mm | 25.4 | 35 |
| Beam aperture | mm | 19.05 | 25 |
| Gap distance | mm | 8.89 | 8.89 |
| Total length | mm | 191 | 211 |
| β | | 1.04 | 1.5 |
| drive power | kW | 10 | 30-50 |
| modulating voltage | kV | 50 | 50 |

A compromised method is adopted in order to decrease the Q factor and limit the wall power loss caused by low Q value at the same time: Two end plates together with the noses on them are made of OFC while other parts of the prebuncher are made of SUS (Stainless Steel). The layout of the prebuncher is shown in Figure 2.

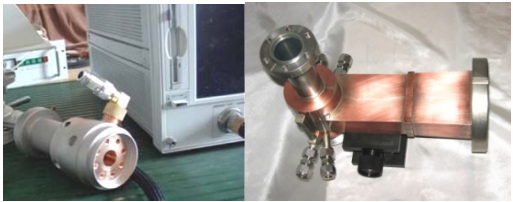

Figure 2. Layout of the prebuncher

## 3 Buncher

The buncher is a copper disk-loaded structure fabricated by brazing in the hydrogen furnace. It is a traveling-wave structure with only six cavities, including the input and output couplers. It is operated in $2\pi/3$ mode with a phase velocity of 0.75 c. A peak field of 7.1MV/m in the fundamental space harmonic with the designed driving power of 3.0 MW is achieved. The parameters and layout of the buncher are shown in detail in Table 3 and Figure 3 respectively.

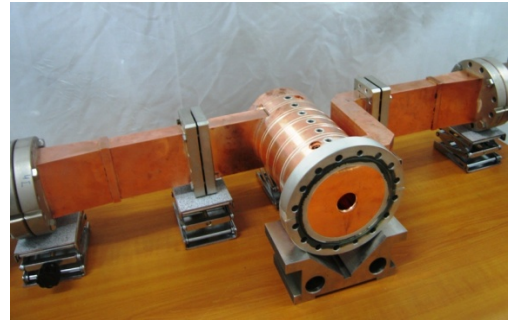

Figure 3. Layout of the buncher

## 4 Accelerating structure

KIPT linac is composed of 10 constant gradient (C.G.) accelerating structures. The structure has got bigger beam aperture than former designs and the 10 C.G. accelerating tubes (Serial number from A0 to A9) which will be used to boost the beam energy to 100 MeV have already been installed in the accelerator tunnel in KIPT (shown in Figure 4). For high intensity electron linacs, the beam break-up (BBU) effect needs to be considered. To suppress the BBU effect, bigger aperture and step were adopted in the structure: the disk hole diameter decreases from 27.887 mm to 23.726 mm in a stepwise fashion along the structure (26.22 mm to 19.093 mm for BEPCII 3m long structure); the average disk hole diameter step increases to ~0.122 mm (~0.085 mm for BEPCII 3m long structure). For further suppressing the BBU effect[3], a method of opening four holes symmetrically distributed on disks is adopted which will get the benefits in improvement of BBU threshold current. According to the simulation, when the diameter of the hole equals to 9 mm (11 mm, 13 mm), the operating frequency ($2\pi/3$ mode and $TM_{01}$ mode) maintains still at 2856 MHz but the $EM_{11}$ wave frequency band will move up by about 6 MHz

Table 3. Parameters of buncher

| Parameter | Unit | Design |
|---|---|---|
| Operating frequency | MHz | 2856 |
| Unloaded $Q_0$ | | 10960 |
| operating mode | | $2\pi/3$ |
| Length | mm | 240.6 |
| cavity diameter (2b) | mm | 83.292 |
| Disk hole diameter (2a) | mm | 22.615 |
| Disk thickness diameter t | mm | 5.84 |
| Number of cells | | 6 |
| VSWR | | 1.06 |
| bandwidth (VSWR<1.2) | MHz | ±2 |
| phase velocity | c | 0.75 |
| Shunt impedance | MΩ/m | 36.3 |
| input power | MW | 2~4 |
| field gradient | MV/m | 5.7~8.1 |

(11 MHz, 16 MHz) with respect to the case without holes in discs. For the 10 C.G. accelerating tubes designed for this project, the four holes are opened in each tube as follows:

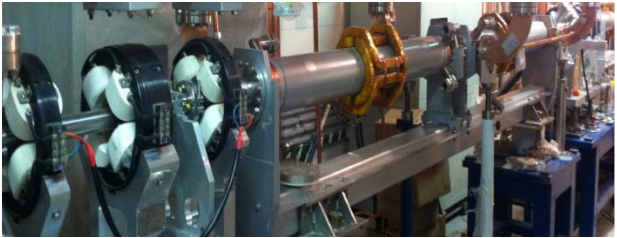

Figure 4. Installed accelerating tubes in the accelerating tunnel in KIPT

Table 4. Measurement results of accelerating tubes

| Serial No. | VSWR (2856 MHz) | | Attenuation constant (dB) | Filling time (ns) |
|---|---|---|---|---|
| | $f_0$ | Frequency bandwidth (MHz) VSWR≤1.2 | | |
| A0 | 1.068 | 5.9 | -1.32 | 221.9 |
| A1 | 1.1 | 5.9 | -1.33 | 221.3 |
| A2 | 1.065 | 5.5 | -1.35 | 221.2 |
| A3 | 1.13 | 5.0 | -1.55 | 222.0 |
| A4 | 1.054 | 6.0 | -1.28 | 217.0 |
| A5 | 1.083 | 6.0 | -1.29 | 224.0 |
| A6 | 1.09 | 5.2 | -1.38 | 222.4 |
| A7 | 1.041 | 5.7 | -1.35 | 220.6 |
| A8 | 1.12 | 5.1 | -1.45 | 221.3 |
| A9 | 1.105 | 5.0 | -1.44 | 223.5 |

For A0 tube, there are no holes.

For A1 (A2, A3) tube, there are 4 holes from the 3rd to the 6th disks with diameter of 9 mm (11 mm, 13 mm) and the distance between the centre of the hole and the centre of the beam aperture is 28mm. The A4/A7, A5/A8 and A6/A9 tubes are the same as the A1, A2 and A3 tubes respectively. The measurement results of the 10 tubes are listed in Table 4.

The methods adopted above for suppressing the BBU effect are feasible theoretically, however, the results of the mentioned methods will be observed in the future testing and commissioning in the accelerating tunnel in KIPT, and we will discuss it in future papers.

## 5 Conclusion

The work of accelerating structure design and fabrication for a project with the participation of NSC KIPT, ANL and IHEP together are described in this paper. All of the prebuncher, buncher as well as the C.G. accelerating tubes have got a satisfying measurement result. The methods for suppressing the BBU effect are considered and adopted because of the high intensity of this electron linac. Now two thirds of the assembling work of the linac was already finished in NSC KIPT, and all the accelerating structures will be tested in the future commissioning stage.